\documentclass{PoS}

\usepackage{mathtools}

\newcommand{\newc}{\newcommand}
\newc{\be}{\begin{equation}}
\newc{\ee}{\end{equation}}
\newc{\beq}{\begin{equation}}
\newc{\eeq}{\end{equation}}
\newc{\bea}{\begin{eqnarray}}
\newc{\eea}{\end{eqnarray}}
\newc{\simlt}{~\mbox{\smaller\(\lesssim\)}~}
\newc{\simgt}{~\mbox{\smaller\(\gtrsim\)}~}

\def\vev#1{\left\langle #1\right\rangle}

\newcommand{\pmatr}[1]{\begin{pmatrix} #1 \end{pmatrix}}

\title{A Simplified Twin Pati-Salam Theory of Flavour with a TeV Scale Vector Leptoquark}

\ShortTitle{Twin Pati-Salam}

\author{\speaker{Stephen~F.~King}\thanks{The author expresses sincere 
thanks to the organisers for the invitation and their hospitality and 
acknowledges the STFC Consolidated Grant ST/L000296/1 and the European Union's Horizon 2020 Research and Innovation programme under Marie Sk\l{}odowska-Curie grant agreement HIDDeN European ITN project (H2020-MSCA-ITN-2019//860881-HIDDeN).}\\
        School of Physics and Astronomy, University of Southampton,
SO17 1BJ Southampton, UK \\
        E-mail: \email{king@soton.ac.uk}}

\abstract{In this talk, we discuss a simplified version of a recently proposed
twin Pati-Salam (PS) theory of flavour broken to the $G_{4321}$ gauge group at high energies, then to the Standard Model at low energies,
yielding a TeV scale vector leptoquark $U^{\mu}_1(3,1,2/3)$ which may be used to address the 
lepton universality anomalies $R_{K^{(*)}}$ and $R_{D^{(*)}}$ in $B$ decays.
The model also accounts for quark and lepton mass hierarchies, providing a link between the flavour anomalies and a theory of flavour.
}

\FullConference{%
  Corfu Summer Institute 2021 "School and Workshops on Elementary Particle Physics and Gravity"\\
  29 August - 9 October 2021\\
  Corfu, Greece
}

\begin{document}

\section{Introduction}
Last year, new evidence was presented for 
the semi-leptonic $B$ decay ratio $R_{K^{(*)}}$, which violates $\mu - e$ universality in $b\rightarrow s\mu \mu$ decays~\cite{Aaij:2021vac}. 
Together with the semi-leptonic $B$ decay ratio $R_{D^{(*)}}$, which violates 
$\tau$ universality in $b\rightarrow c \tau \nu_{\tau}$ decays, these anomalies motivate new theories of flavour involving leptoquarks.
Indeed, the single vector leptoquark $U^{\mu}_1(3,1,2/3)$ has been shown to address all the B physics anomalies 
\cite{Alonso:2015sja,Calibbi:2015kma,Barbieri:2015yvd,Sahoo:2016pet,Buttazzo:2017ixm,Assad:2017iib,Barbieri:2017tuq,Kumar:2018kmr,Crivellin:2018gyy,Cornella:2019hct,Crivellin:2019szf,Dev:2020qet,Fuentes-Martin:2019ign,Fuentes-Martin:2020luw,Fuentes-Martin:2020hvc,Bhaskar:2021pml,Iguro:2021kdw,Angelescu:2021lln,Cornella:2021sby,Hiller:2021pul}.

Such a vector leptoquark is predicted by Pati-Salam theory  (PS) \cite{Pati:1974yy}, but its mass has to be too heavy to explain the anomalies, in order to not violate certain experimental limits. Notwithstanding this, 
such a vector leptoquark could be made sufficiently light (around the TeV scale) consistent with the experimental limits,
\footnote{A low energy PS gauge group has also been considered from a different perspective \cite{Perez:2013osa}.}
in more complicated versions of the PS theory~\cite{DiLuzio:2017vat,Fornal:2018dqn,Baker:2019sli,Calibbi:2017qbu,Bordone:2017bld,Heeck:2018ntp,Matsuzaki:2018jui,Blanke:2018sro}.
However, such theories are typically defined as effective theories, and 
their ultraviolet completion remains obscure,which 
motivates further model building. 
Of particular interest are those models which might also account for the origin of quark and lepton masses, if such theories could be constructed.
In this case, the $B$ physics anomalies could be the first indication of a new theory of flavour beyond the Standard Model (BSM).

\begin{figure}[ht]
\centering
	\includegraphics[scale=0.23]{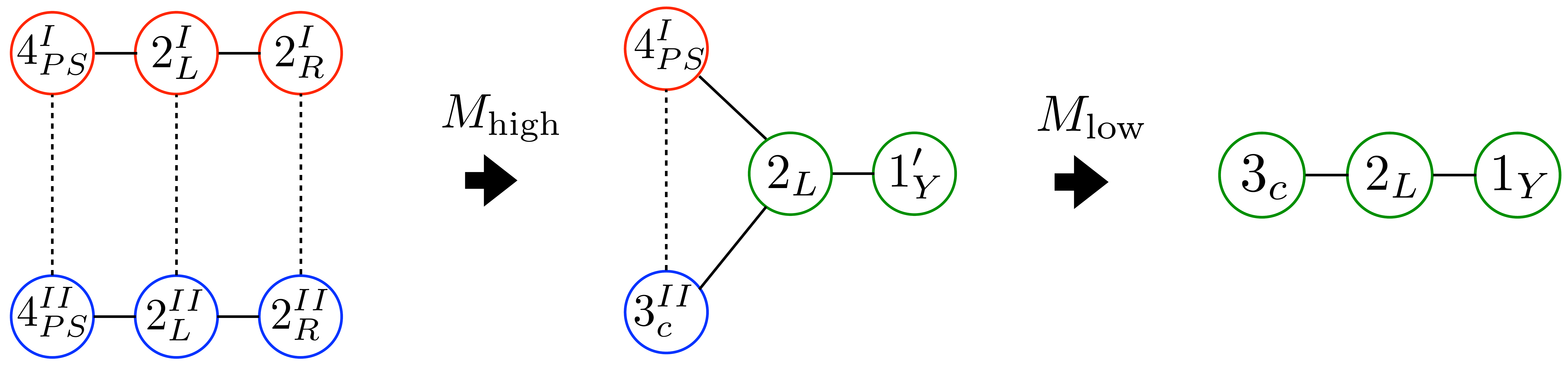}
\caption{The model is based on two copies of the PS gauge group $G_{422}=SU(4)_{PS}\times SU(2)_L\times SU(2)_R$. 
The circles represent the gauge groups with the indicated symmetry breaking as in Eq.\ref{breaking}, ending with the 
Standard Model (SM) gauge group $G_{321}=SU(3)_{c}\times SU(2)_L\times U(1)_Y$. }
\label{model}
\end{figure}

In a recent paper \cite{King:2021jeo} we proposed such a theory of flavour, 
not only capable of explaining some of the anomalies, for natural values of the parameters, but also providing an explanation of quark and lepton (including neutrino) mass and mixing hierarchies. The basic premise is that, at high energies, the theory involves two copies of the PS gauge group, $G_{422}$ \cite{Pati:1974yy}. Having two copies of the PS groups allows gauge boson masses at different mass scales, as follows.

The twin PS gauge groups are assumed to be broken in stages first to 
$G_{4321}$ then to the SM gauge group
$G_{321}$, as depicted in Fig.~\ref{model},
\begin{align}
G_{422}^I \times G_{422}^{II} \stackrel{M_{\mathrm{high}}}{\longrightarrow} G_{4321}
\stackrel{M_{\mathrm{low}}}{\longrightarrow} G_{321}
\label{breaking}
\end{align}
The high scale PS symmetry group $SU(4)^{II}_{PS}$ is broken at $M_{\mathrm{high}} \gtrsim 1$ PeV, the latter limit being due to the non-observation of $K_L\rightarrow \mu e$
\cite{Valencia:1994cj},
where $M_{\mathrm{high}}$ may be as high as the conventional scale of Grand Unified Theories (GUTs).
The low scale PS group $SU(4)^{I}_{PS}$ is broken at $M_{\mathrm{low}} \sim 1$ TeV in order to explain the anomalies.

We assume that three chiral fermion families transform under the second PS group, 
$G_{422}^{II}$, while a fourth vector-like (VL) fermion family, transforms under the other the first PS group, $G_{422}^{I}$. 
The mixing of the VL fermions with the chiral quarks and leptons will play a role in controlling 
the couplings of ordinary quarks and leptons to the PS gauge bosons, which will emerge as mixed states.
In the absence of such mixing, ordinary quarks and leptons would not couple at all to the low scale 
vector leptoquarks, and would be massless due to the absence of SM Higgs doublets.

The explanation of the anomalies involves the low scale vector leptoquark $U^{\mu}_1(3,1,2/3)$ from the $SU(4)^{I}_{PS}$,
broken at $M_{\mathrm{low}} \sim 1$ TeV, which couples to only the VL fermions initially, but which will also couple to ordinary quarks and leptons via their mixing with the VL fermions. The controlled nature of these couplings will ensure that the experimental limits are respected.
Similarly, the origin of quark and lepton masses also depends on the mixing with the VL fermions, and will involve also new
 ``personal'' Higgs doublets for the second and third family fermion masses, where the origin and nature of these fields is very different from the ``private'' Higgs doublets envisaged in~\cite{Porto:2007ed,Porto:2008hb,BenTov:2012cx,Rodejohann:2019izm}.

Since the explanation of the anomalies and the origin of quark and lepton masses both arise from mixing with the VL fermions,
this provides a link between the flavour anomalies and the theory of flavour.
More details about the simplified model are given in the next section.

\section{A Simplified Twin Pati-Salam Theory of Flavour}
\label{high}
In this section, we shall consider a simplified version of the full twin PS model \cite{King:2021jeo} in which we address the question of the second and third family quark and lepton masses only. In this simplified version, the first family is assumed to be massless, along with the neutrino masses,
both of which are not directly related to the anomalies. In this simplified version of the model, presented in the Corfu talk, 
only a single vector-like family is required, and the model is easier to formulate.
For details of the full model, see~\cite{King:2021jeo}.

\subsection{The High Energy Model}
\label{2.1}
It is well known that quarks and leptons may be unified into the traditional Pati-Salam (PS) gauge group \cite{Pati:1974yy},
\be
G_{422}=SU(4)_{PS}\times SU(2)_L\times SU(2)_R
\label{PS}
\ee
In the traditional PS theory, the left-handed (LH) chiral quarks and leptons are unified into 
$SU(4)_{PS}$ multiplets with leptons as the fourth colour (red, blue, green, lepton),
\begin{equation}
{\psi_i}(4,2,1)=
\left(\begin{array}{cccc}
u_r & u_b & u_g & \nu \\ d_r & d_b & d_g & e^-
\end{array} \right)_i \equiv (Q_i,L_i)
\label{psi}
\end{equation}
\begin{equation}
\psi^c_j(\bar{4},1,\bar{2})=
\left(\begin{array}{cccc}
u^c_r & u^c_b & u^c_g & \nu^c \\ d^c_r & d^c_b & d^c_g & e^c
\end{array} \right)_j \equiv(u^c_j, d^c_j, \nu^c_j, e^c_j)
\label{psic}
\end{equation}
where $\psi^c_j$ are the CP conjugated RH quarks and leptons (so that they become LH) forming $SU(2)_R$ doublets
and 
$i,j=1\ldots 3$ are family indices. Three right-handed neutrinos (actually their CP conjugates $\nu^c_j$) are 
predicted as part of the gauge multiplets. 

In the twin PS model in Table~\ref{twinPS}, 
the usual three chiral fermion families originate from the second PS group $G_{422}^{II}$, broken at the high scale.
There are no standard scalar Higgs doublet fields 
which transform as $({\bf 1},{\overline{\bf 2}},{\bf 2})$ under $G_{422}^{II}$, hence no standard Yukawa couplings involving the chiral fermions. These will be generated effectively via mixing with the vector-like (VL) fermions which only have quantum numbers under the first PS group,
$G_{422}^{I}$. This mixing is facilitated by the non-standard Higgs scalar doublets contained in $\phi,\overline{\phi}, H,\overline{H}$ in 
Table~\ref{twinPS}, via the couplings,
\begin{align}
{\cal L}^{ren}_4 = 
& y^{\psi}_{i4}\overline{H}  \psi_i {\psi^c_4} 
+  y^{\psi}_{43}{H} {\psi_4} \psi^c_3
+x^{\psi}_{i4}{\phi} \psi_i \overline{\psi_4} 
+ x^{\psi^c}_{4j} \overline{\psi^c_4} \overline{\phi} \psi^c_j
+ M^{\psi}_{4}\psi_4 \overline{\psi_4}
+ M^{\psi^c}_{4}\psi^c_4 \overline{\psi^c_4} \label{L4}
\end{align}
where $i,j=2,3$, and 
$x,y$ are dimensionless coupling constants and $M_4$ are the VL masses.
These couplings mix the chiral fermions with the VL fermions, and 
will be responsible for generating effective Yukawa couplings of the second and third families.

We shall refer to the Higgs doublets contained in $H,\overline{H}$ as personal Higgs doublets, since under the Standard Model decomposition, there will be a separate Higgs for each fermion mass as we shall see shortly.
The Higgs singlet fields in $\phi,\overline{\phi}$ are called Yukons, since they are necessary to generate the effective Yukawa couplings.

Since the VL fermions will mix with the second and third chiral families, they lead to effective couplings to TeV scale
$SU(4)^{I}$ gauge bosons which violate lepton universality between the second and third families.

\begin{table}
\centering
\begin{tabular}{| l | c  c c | c c c| }
\hline
Field & $SU(4)^I_{PS}$ & $SU(2)^{I}_L$ & $SU(2)^{I}_R$ &  $SU(4)^{II}_{PS}$ & $SU(2)^{II}_L$ & $SU(2)^{II}_R$ \\ 
\hline \hline
$\psi_{1,2,3}$ 		 & ${\bf 1}$  & ${\bf 1}$  & ${\bf 1}$ & ${\bf 4}$ & ${\bf 2}$ & ${\bf 1}$  \\
$\psi^c_{1,2,3}$ 	 & ${\bf 1}$  & ${\bf 1}$ & ${\bf 1}$ & ${\overline{\bf 4}}$ & ${\bf 1}$ & ${\overline{\bf 2}}$\\
\hline
$\psi_{4}$ 		& ${\bf 4}$ & ${\bf 2}$ & ${\bf 1}$ & ${\bf 1}$ & ${\bf 1}$ & ${\bf 1}$  \\
$\overline{\psi_{4}}$ 		& ${\overline{\bf 4}}$   & ${\overline{\bf 2}}$ & ${\bf 1}$ & ${\bf 1}$ & ${\bf 1}$ & ${\bf 1}$ \\
$\psi^c_{4}$ 		 & ${\overline{\bf 4}}$ & ${\bf 1}$  & ${\overline{\bf 2}}$ & ${\bf 1}$ & ${\bf 1}$ & ${\bf 1}$\\
$\overline{\psi^c_{4}}$ 		& ${\bf 4}$  & ${\bf 1}$  & ${\bf 2}$ & ${\bf 1}$ & ${\bf 1}$ & ${\bf 1}$ \\
\hline
$\phi$  &   ${\bf 4}$  & ${\bf 2}$ &  ${\bf 1}$ & ${\overline{\bf 4}}$ & ${\overline{\bf 2}}$ &  ${\bf 1}$  \\
$\overline{\phi}$ &   ${\overline{\bf 4}}$ & ${\bf 1}$ & ${\overline{\bf 2}}$ & ${\bf 4}$ &  ${\bf 1}$ & ${\bf 2}$  \\
\hline
\hline
 $H$   & ${\overline{\bf 4}}$ & ${\overline{\bf 2}}$ &  ${\bf 1}$ &  ${\bf 4}$   &  ${\bf 1}$  &  ${\bf 2}$  \\
$\overline{H}$   & ${\bf 4}$ &  ${\bf 1}$  &  ${\bf 2}$  & ${\overline{\bf 4}}$  &  ${\overline{\bf 2}}$ & ${\bf 1}$  \\
\hline
\hline
\end{tabular}
\caption{The twin PS theory based on 
$G_{422}^I \times G_{422}^{II}$.  The model consists of 
three left-handed chiral fermion families $\psi_{1,2,3},\psi^c_{1,2,3}$ under the second PS group,
plus a VL fourth fermion family $\psi_{4},\psi^c_{4}$ and their conjugates under the first PS group. 
Personal Higgs doublets are contained in $H,\overline{H}$, one for each fermion.
The Higgs singlets in $\phi, \overline{\phi}$ are called Yukons.
In addition to the fields shown here,
we require further high energy Higgs fields (not shown) whose VEVs will break the second PS group at a high scale, leaving the first unbroken.
We also need further Higgs fields, which break the two left-right gauge groups into their diagonal subgroup.
}
\label{twinPS}
\end{table}

\subsection{Effective Yukawa operators}
\label{2.2}

We have already remarked that the usual Yukawa couplings involving purely chiral fermions 
are absent. In this subsection we show how they may be generated effectively once the vector-like fermions are integrated out.

With a fourth VL family, we may write the masses and couplings in Eq.\ref{L4} as a $5\times 5$ matrix in flavour space
\be
	M^{\psi} = \pmatr{
	&\psi^c_1&\psi^c_2&\psi^c_3& \overline{\psi_4} &\psi^c_4\\ 
	\hline
	\psi_1|&0&0&0&0 &0\\
	\psi_2|&0&0&0&0&y^{\psi}_{24}\overline{H} \\
	\psi_3|&0&0&0&x^{\psi}_{34} {\phi}&y^{\psi}_{34}\overline{H} \\ 
	\psi_4|&0& 0 & y^{\psi}_{43} {H}&M^{\psi}_{4}&0\\ 
	\overline{\psi^c_4}|&0&x^{\psi^c}_{42}\overline{\phi}&x^{\psi^c}_{43}\overline{\phi}&0& M^{\psi^c}_{4}}.	
	\label{M^psi_an_1}
\ee
where the extra zeroes are achieved by $(\psi_2,\psi_3)$ rotations, where such rotations leave the upper $3\times 3$ block of zeroes unchanged, so 
the form of Eq.\ref{M^psi_an_1} is just a convenient choice of basis.

\begin{figure}[ht]
\centering
	\includegraphics[scale=0.1]{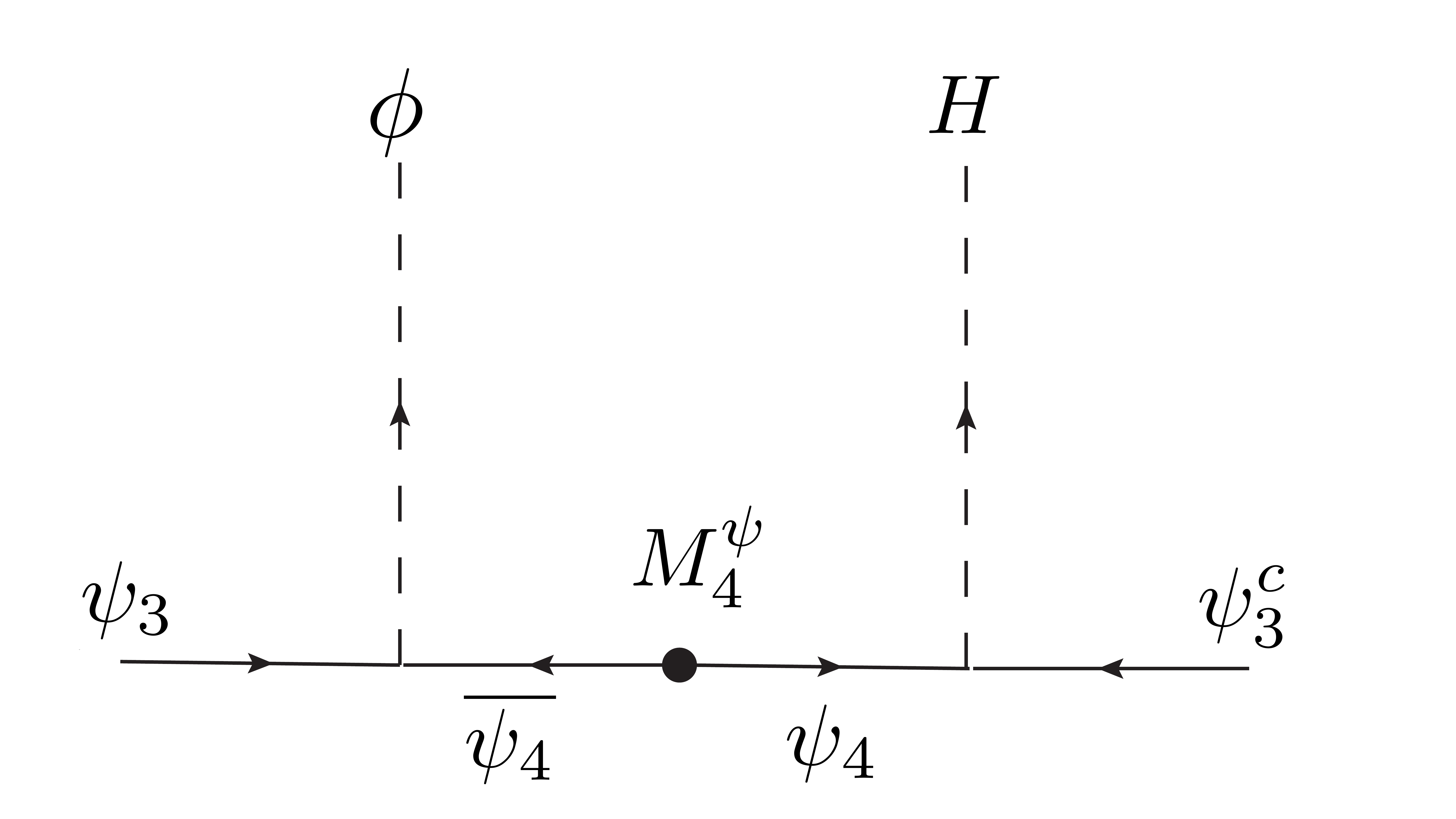}
\hspace*{1ex}
	\includegraphics[scale=0.1]{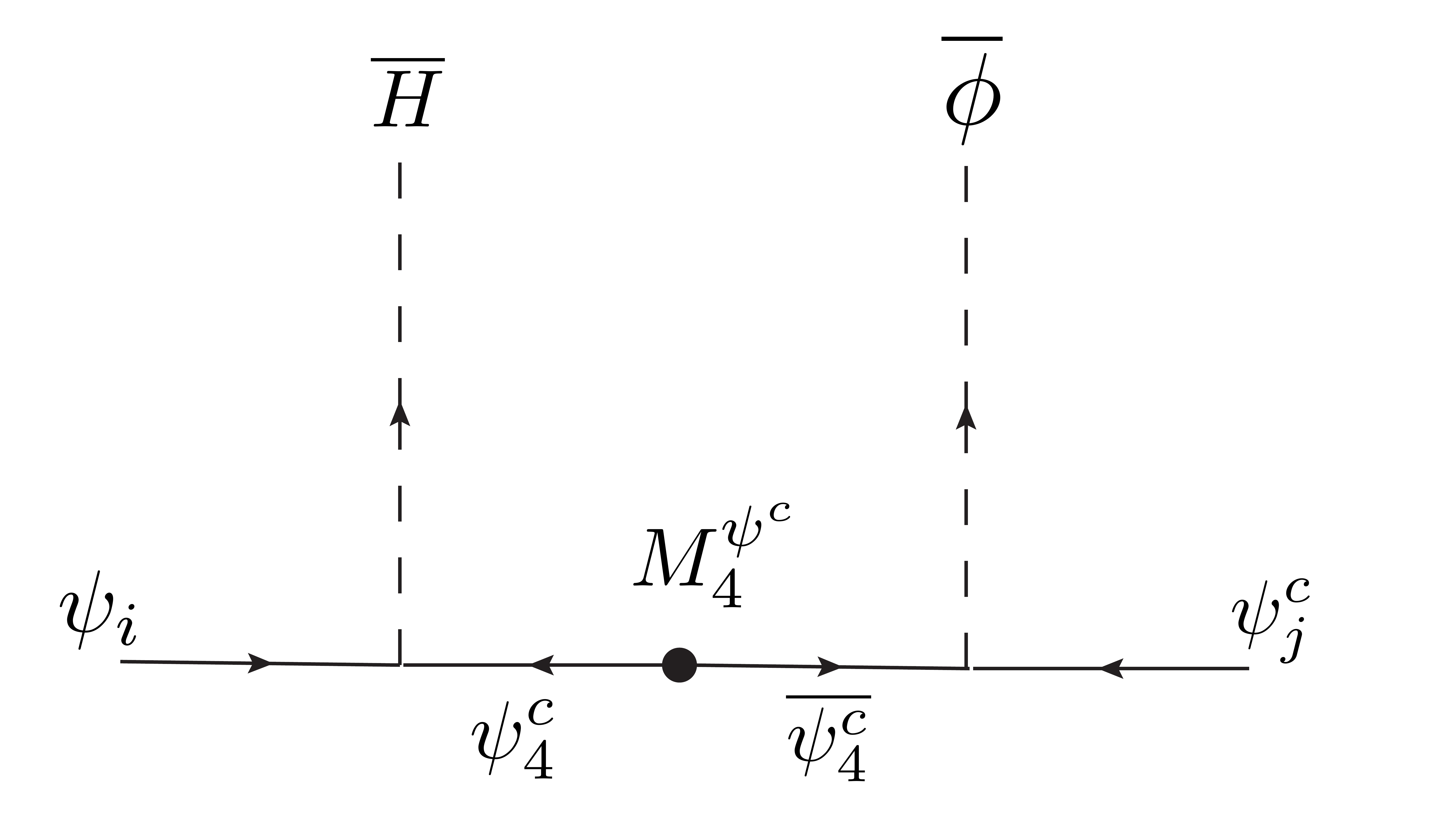}
\caption{Diagrams which lead to the effective Yukawa couplings of the third family (left panel) and second family (right panel) where $i,j=2,3$ are the only non-zero values. }
\label{Fig1}
\end{figure}

There are several distinct mass scales in this matrix: the personal Higgs VEVs $\langle H \rangle$, 
$\langle \overline{H} \rangle$, the Yukon VEVs $\langle \phi \rangle$, $\langle  \overline{\phi} \rangle$
and the VL fourth family masses $M^{\psi}_{4}$, $M^{\psi^c}_{4}$. 
Assuming the latter are heavier than all the Higgs VEVs, we may integrate out the fourth family, to generate 
effective Yukawa couplings of the quarks and leptons which originate from the diagrams in 
Fig.~\ref{Fig1}. 

The two diagrams in Fig.\ref{Fig1} lead to effective Yukawa operators (up to an irrelevant minus sign),
after integrating out VL fermions,
\be
{\cal L}^{Yuk}_{4eff}= \frac{x^{\psi}_{34} {\phi}  y^{\psi}_{43} {H} }{M^{\psi}_{4}}  \psi_3 {\psi^c_3}
+
\frac{ y^{\psi}_{i4} \overline{H}  x^{\psi^c}_{4j}  \overline{\phi}  }{M^{\psi^c}_{4}} \psi_i {\psi^c_j}+H.c.
\label{Yuk_mass_insertion4}
\ee
where $i,j=2,3$.
After Pati-Salam breaking, 
these terms will lead to Yukawa matrices for quarks and leptons as we now discuss.

\begin{figure}[ht]
\centering
	\includegraphics[scale=0.25]{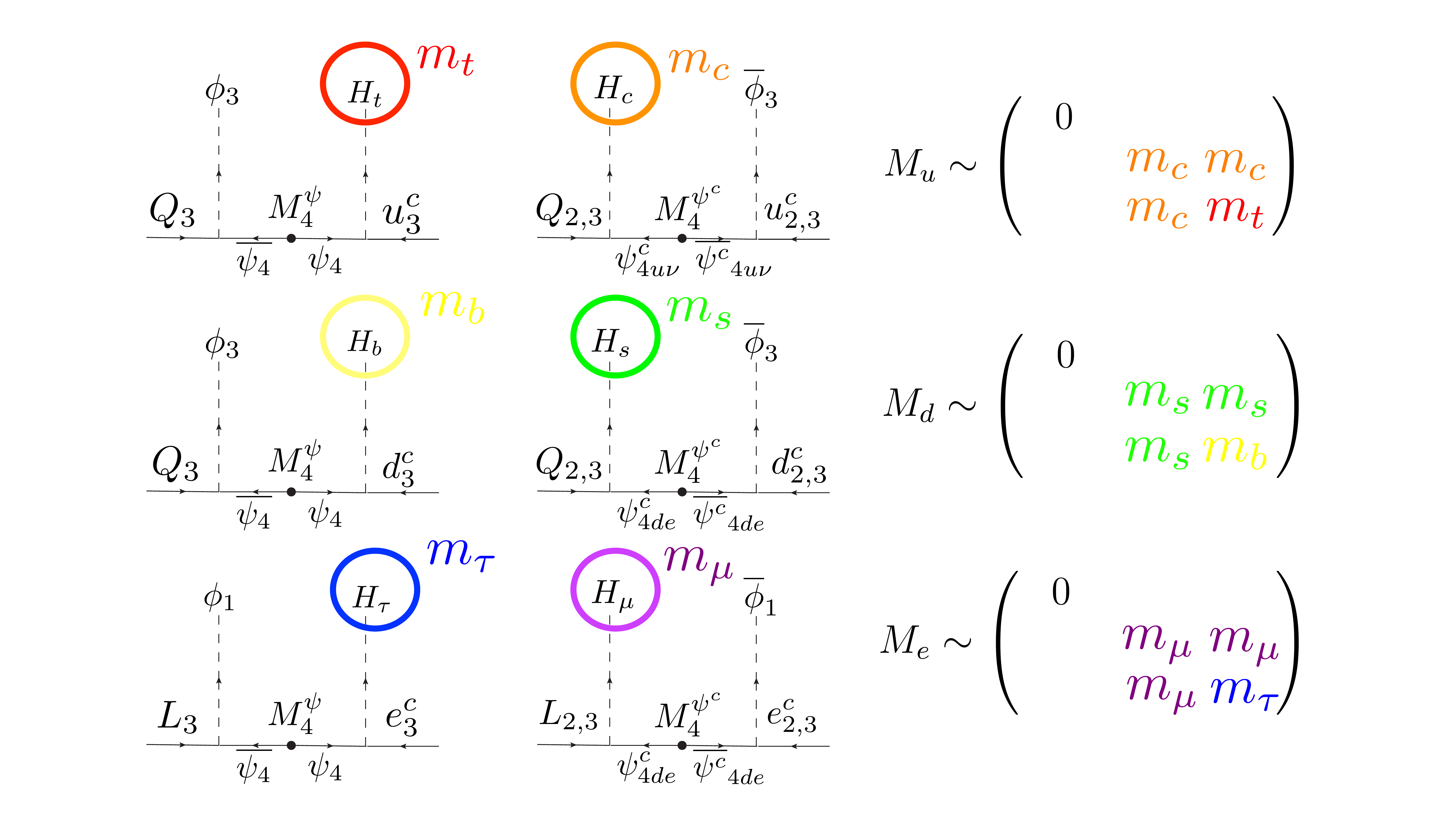}
\caption{Diagrams which lead to the second and third family quark and charged lepton mass matrices of the SM.
These diagrams originate from those shown in Fig.~\ref{Fig1}, when decomposed under the SM gauge group.
As indicated by the colour coding, there is a separate personal Higgs doublet responsible for each of the second and third family quark and lepton masses, which originates from the decomposition of the scalars $H,\overline{H}$ in Table~\ref{twinPS} under the SM gauge group.
The Higgs singlet fields from the decomposition of the scalars $\phi,\overline{\phi}$ are called Yukons, since they are necessary to generate the effective Yukawa couplings as shown in the diagram.}
\label{Fig2}
\end{figure}

The reason we have gone to the basis in Eq.\ref{M^psi_an_1}, 
with more zeros in the $\psi_4,\overline{\psi_4}$ entries, is that the effective Yukawa operators in Eq.\ref{Yuk_mass_insertion4}
have the suggestive matrix form,
\be
{\cal L}^{Yuk}_{4eff}= 
\pmatr{
	&\psi^c_1&\psi^c_2&\psi^c_3\\ 
	\hline
	\psi_1|&0&0&0\\
	\psi_2|&0&0&0\\
	\psi_3|&0&0&x^{\psi}_{34} y^{\psi}_{43}}
	\frac{{\phi} }{M^{\psi}_{4}}  {H} 
	+
\pmatr{
	&\psi^c_1&\psi^c_2&\psi^c_3\\ 
	\hline
	\psi_1|&0&0&0\\
	\psi_2|&0 & y^{\psi}_{24} x^{\psi^c}_{42} & y^{\psi}_{24} x^{\psi^c}_{43}\\
	\psi_3|&0 & y^{\psi}_{34} x^{\psi^c}_{42} & y^{\psi}_{34} x^{\psi^c}_{43}}
	\frac{\overline{\phi} }{M^{\psi^c}_{4}}\overline{H}  
\label{Yuk_mass_insertion_1}
\ee
where the dimensionless couplings $x,y$ in the matrices are expected to be of order unity.
If we assume that $\phi,\overline{\phi}$ fields develop vacuum expectation values (VEVs)
with a hierarchy of scales,
\be
\frac{\langle \overline{\phi} \rangle}{M^{\psi^c}_{4}}\ll \frac{\langle  {\phi} \rangle}{M^{\psi}_{4}}  \lesssim 1
\label{hierarchy}
\ee
then the first matrix in Eq.\ref{Yuk_mass_insertion_1} 
generates larger effective third family Yukawa couplings, while the second matrix generates suppressed second
family Yukawa couplings and mixings.
Since the sum of the two matrices has rank 1, the first family will be massless.
When decomposed under the SM gauge group, the operators in Eq.\ref{Yuk_mass_insertion_1} lead to the 
second and third family quark and charged lepton mass matrices as shown in 
Fig.~\ref{Fig2}.

\section{Phenomenology of $G_{4321}$ }

In this section we shall discuss the phenomenology resulting from the 
low scale symmetry breaking
\begin{align}
 G_{4321} \stackrel{M_{low}}{\longrightarrow} G_{321}
 \label{4321to321}
\end{align}
which is achieved from 
the low scale VEVs,
\beq  
\label{vevconf}
\vev{\phi_3} = 
\left(
\begin{array}{ccc}
\tfrac{v_3}{\sqrt{2}} & 0 & 0 \\
0 & \tfrac{v_3}{\sqrt{2}} & 0 \\ 
0 & 0 & \tfrac{v_3}{\sqrt{2}} \\
0 & 0 & 0
\end{array}
\right) \, , \ \ 
\vev{\phi_1} = 
\left(
\begin{array}{c}
0 \\ 
0 \\ 
0 \\
\tfrac{v_1}{\sqrt{2}}
\end{array}
\right) \, ,
\eeq
where 
\begin{equation}
v_1 \sim v_3  \lesssim 1 \ {\rm TeV}. \ \ 
\label{HVEV}
\end{equation}

The low scale mass gauge bosons resulting from the symmetry breaking in Eq.~\ref{4321to321} include the following TeV scale gauge bosons:
a massive vector leptoquark $U_1^{\mu} = (3,1,2/3)$,
a massive gluon $g'_{\mu} = (8,1,0)$ 
and a massive $Z'_{\mu} = (1,1,0)$.  

The massive gauge bosons $U_1,g',Z'$ couple to the chiral fermions 
and VL fourth family fermions with left-handed interactions~\cite{DiLuzio:2017vat},
\begin{align}
& \frac{g_4}{\sqrt{2}}  \left( \bar{Q}_{L4} \gamma^\mu L_{L4} + \textrm{H.c.} \right) U_{1\mu} \nonumber \\
 + & \frac{g_4 g_s}{g_3} \left( \bar{Q}_{L4} \gamma^\mu T^a Q_{L4} - \frac{g_3^2}{g_4^2} \,
\bar{Q}_{Li} \gamma^\mu T^a Q_{Li} \right) g'^a_\mu \label{L} \\
+\frac{\sqrt{3} \,g_4 g_Y}{\sqrt{2} \,g_1} & \left( \frac{1}{6} \bar{Q}_{L4} \gamma^\mu Q_{L4} -\frac{1}{2} \bar{L}_{L4} \gamma^\mu L_{L4} 
- \frac{g_1^2}{9 g_4^2} \,\bar{Q}_{Li} \gamma^\mu Q_{Li}
+ \frac{ g_1^2}{3 g_4^2} \,\bar{L}_{Li} \gamma^\mu 
L_{Li} \right) Z'_\mu \nonumber
\end{align}
A typical benchmark point is~\cite{DiLuzio:2017vat}:
$g_4 \approx 3$, $g_3 \approx g_s \approx 1$,
$g_1 \approx g_Y \approx 0.36$, 
$M_{Z'} \approx  1.4$~TeV, $M_{U_1} \approx 1.6$~TeV, and $M_{g'} \approx  2.0$~TeV.
This set of parameters has the typical feature that $g_4\gg g_3,g_1$ so that the heavy gauge bosons $g',Z'$ have suppressed couplings to 
light quarks and leptons, according to Eq.\ref{L}, which will inhibit the direct production of these states at the LHC. 
The vector leptoquark with mass $M_{U_1}\approx 1.6$ TeV is below the sensitivity of LHC searches, for couplings 
consistent with the global fit to the B physics anomalies \cite{Buttazzo:2017ixm}.

The key feature of the gauge boson couplings in Eq.\ref{L}
is that, while the massive gluon $g'_{\mu}$ and the $Z'_{\mu}$ couple to all 
chiral and VL quarks and leptons, the vector leptoquark $U_1^{\mu}$ only couples to the fourth family VL fermions.
The reason is that $U_1^{\mu}$ originates entirely from $SU(4)_{I}$, which remains unbroken to low scales, and 
under which the chiral quarks and leptons are singlets.

\begin{figure}[ht]
\centering
	\includegraphics[scale=0.1]{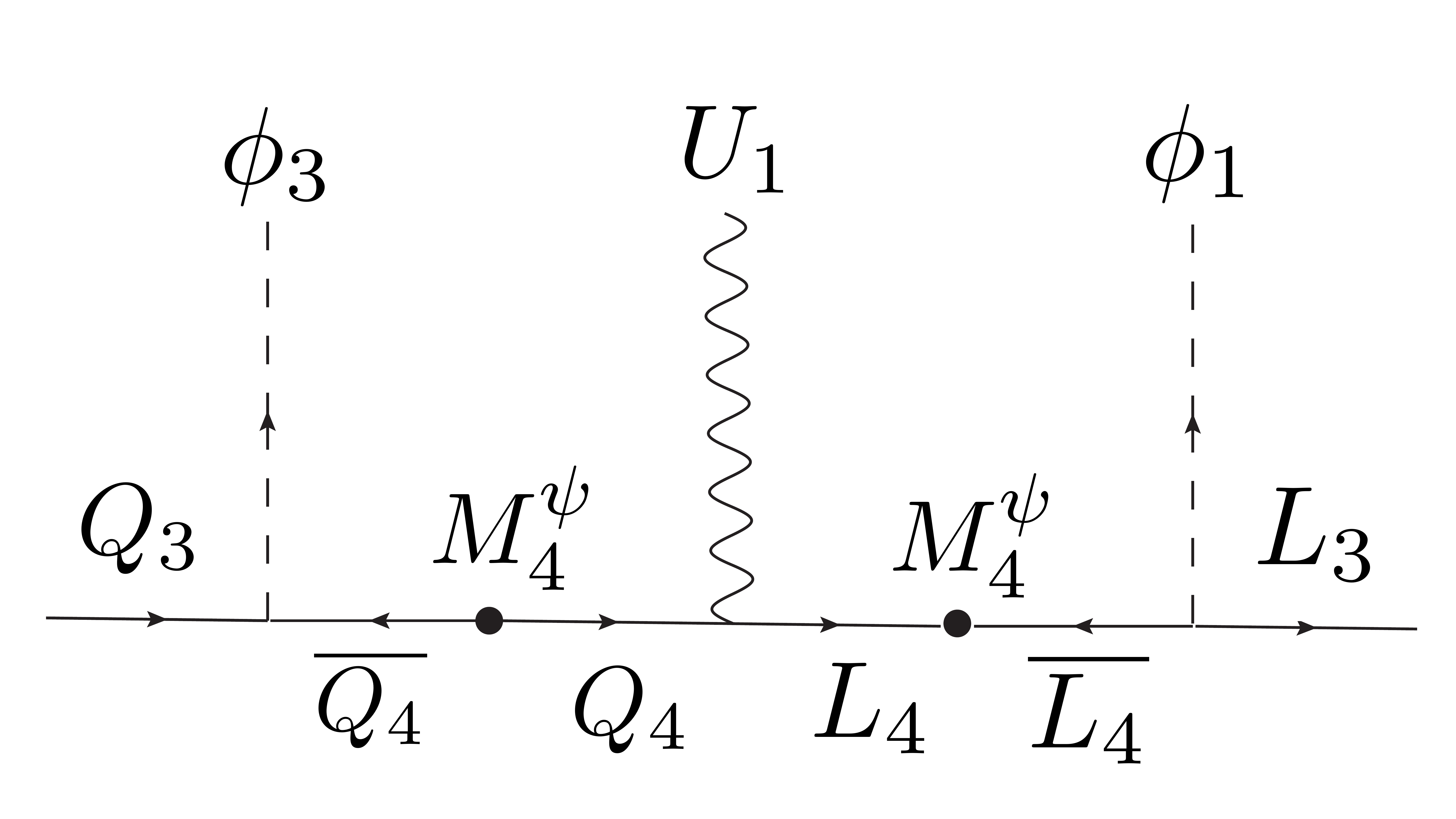}
	\caption{Diagram in the model which leads to the effective $U_1$ vector leptoquark couplings 
to third family quarks to leptons in the mass insertion approximation.  }
\label{Fig4}
\end{figure}

However, effective $U_1^{\mu}$ vector leptoquark couplings to third family quarks and leptons are generated from 
the first line of Eq.\ref{L},
after mixing of the third family chiral fermions with the fourth family VL fermions, leading to the effective operator,
\begin{align}
 \frac{g_4}{\sqrt{2}} 
 \frac{x^{\psi}_{34} {\langle \phi_1 \rangle } }{M^{\psi}_{4}} \frac{x^{\psi}_{34} {\langle \phi_3 \rangle } }{M^{\psi}_{4}}
\bar{Q}_{L3} \gamma^\mu L_{L3} \, U_{1\mu} + H.c.
\label{U1op}
\end{align}
as shown 
in Fig.~\ref{Fig4}.

\begin{figure}[ht]
\centering
	\includegraphics[scale=0.15]{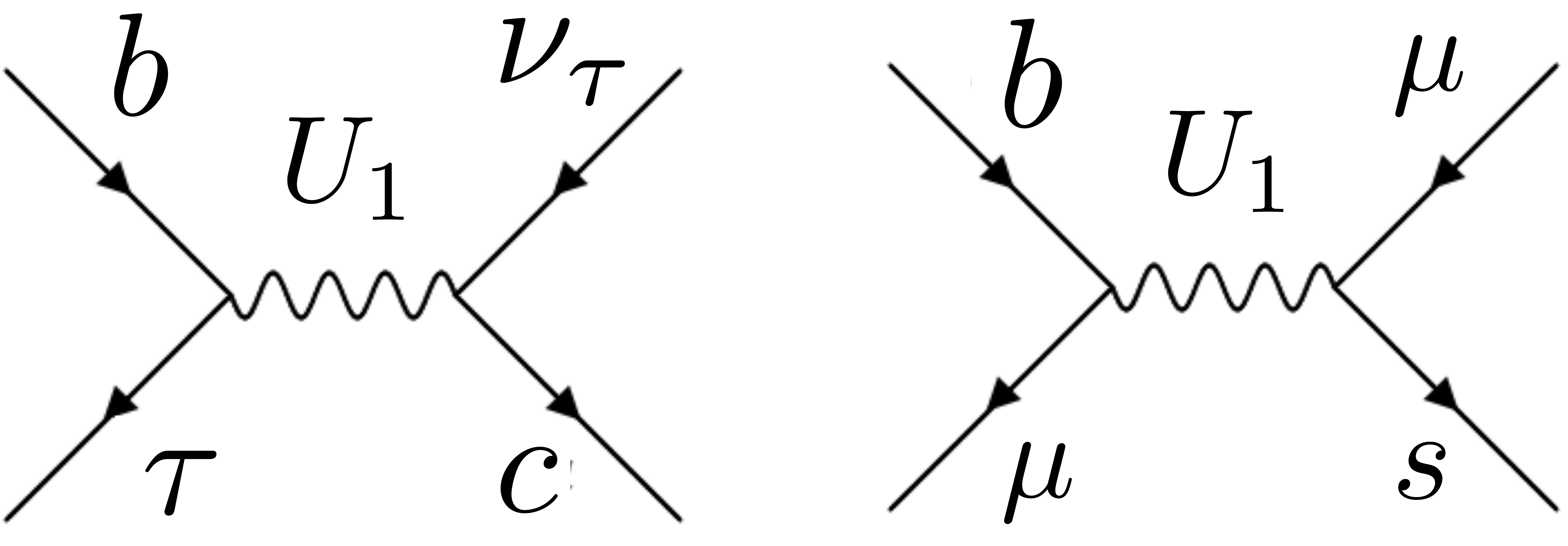}
	\caption{Exchange of the vector leptoquark leads to contributions to $R_{D^{(*)}}$ 
which violates $\tau$ universality in $b\rightarrow c \tau \nu_{\tau}$ decays (left), and $R_{K^{(*)}}$ 
which which violates $\mu - e$ universality in $b\rightarrow s\mu \mu$ decays (right).  }
\label{FigR}
\end{figure}

The operator in Eq.\ref{U1op} has the right structure of vector leptoquark $U_1^{\mu}$ couplings to account for the $B$-physics anomalies
in  $R_{K^{(*)}}$ and $R_{D^{(*)}}$ as discussed in many papers mentioned in the Introduction.
For example, according to the analysis in \cite{Buttazzo:2017ixm}, a single operator as in Eq.\ref{U1op}, involving only the third family doublets,
can account for both the anomalies simultaneously, once the further transformations required to diagonalise the quark and lepton mass matrices are taken into account, leading to, in the notation of \cite{Buttazzo:2017ixm},
\begin{align}
\frac{g_4}{\sqrt{2}} s^Q_{34} s^L_{34} 
\bar{Q}_{L3} \gamma^\mu L_{L3} \, U_{1\mu} \equiv
g_U
\bar{Q}_{L3} \gamma^\mu L_{L3} \, U_{1\mu}  \rightarrow 
g_U \beta_{i\alpha}
\bar{Q}_{Li} \gamma^\mu L_{L\alpha} \, U_{1\mu} 
\label{U1op2}
\end{align}
The couplings in Eq.\ref{U1op2} give rise to the diagrams in Fig.\ref{FigR} in which the exchange of the vector leptoquark $U_{1}$
leads to contributions to lepton flavour non-universality in $R_{D^{(*)}}$ and $R_{K^{(*)}}$.
In the effective field theory analysis of \cite{Buttazzo:2017ixm} these further transformations were regarded as 
relatively free parameters with good global fits to $R_{D^{(*)}}$ and $R_{K^{(*)}}$ obtained for 
$\beta_{s\tau}\approx 4|V_{cb}|$, with $\beta_{b\mu} < 0.5$ and $\beta_{s\mu}<  5|V_{cb}|$
constrained to lie on narrow contours \cite{Buttazzo:2017ixm}. However in the present model 
the quark and lepton mass matrices are predicted, as in Fig.~\ref{Fig2},
and the natural expectation is that these mixing 
parameters are of order $|V_{cb}|$, so it does not seem possible to account for $R_{D^{(*)}}$.

\section{Conclusion}
We have shown that the flavour puzzle, namely the origin of fermion mass and mixing, may be related to recent B physics anomalies.
It is well known that the $R_{K^{(*)}}$ and $R_{D^{(*)}}$ anomalies, related to lepton flavour universality violation,
may be simultaneously explained by a TeV scale vector leptoquark $U_1$ 
of the kind found in Pati-Salam theories.
However the traditional Pati-Salam gauge group must be broken above the PeV scale to avoid the experimental
non-observation of $K_L\rightarrow \mu e$, which is at odds with the requirement of a TeV scale vector leptoquark.
In order to reconcile these conflicting requirements 
requires some non-trivial UV completion of the Pati-Salam theory.
The twin PS model discussed here achieves this, while also accounting for 
quark and lepton masses and mixing.
The quark and lepton mass spectrum, and their couplings to the vector leptoquark, both arise from mixing with the VL fermions,
providing a link between the flavour anomalies and the theory of flavour.
Unfortunately, we have seen that the predicted mass matrices do not yield large enough couplings to explain $R_{D^{(*)}}$, given the current data.
However, it may be possible to account for $R_{D^{(*)}}$ and $R_{K^{(*)}}$, while satisfying all flavour changing and collider constraints, in a 
variation of this model, and work is in progress to investigate this, along the lines of the analysis in \cite{Navarro:2021sfb}.

\end{document}